\documentclass[prc,superscriptaddress,showpacs,floatfix]{revtex4}

\usepackage[dvips]{graphicx}
\usepackage{epsfig}
\usepackage{pst-plot}
\usepackage{bm}
\usepackage{mathrsfs}
\usepackage{amsfonts}
\usepackage{amssymb}
\usepackage{lscape}

\begin{document}

\title[Resonances in deformed meanfields]
{Study of Resonant Structures in a Deformed Mean-Field by 
the Contour Deformation Method in Momentum Space.}
\author{G.~Hagen}
\affiliation{Department of Physics and Technology, University of Bergen, \\
N-5007 Bergen, Norway}  
\author{J.~S.~Vaagen}
\affiliation{Department of Physics and Technology, University of Bergen, \\
N-5007 Bergen, Norway}
\date{\today}

\begin{abstract}
Solution of the momentum space Schr\"odinger equation 
in the case of deformed fields is being addressed. In particular it is shown that a complete set 
of single particle states which includes bound, resonant and complex continuum states may be 
obtained by the Contour Deformation Method. This generalized basis in the complex energy plane is known 
as a Berggren basis. The momentum space Schr\"odinger equation is an integral equation which is easily 
solved by matrix diagonalization routines even for the case of deformed fields. The method is demonstrated for axial symmetry and a fictitious "deformed $^5$He", but may be extended to more
general deformation and applied to truly deformed halo nuclei.
\end{abstract}

\maketitle

\section{Introduction}
\label{sec:introduction}

In nuclear physics, like in atomic physics, expansion of many-body wavefunctions on
single particle bases, generated by a suitable potential has been 
common practice. 
The newly developed Gamow Shell Model 
\cite{michel1, michel2, michel3, witek1, witek2, roberto, betan, betan2,hagen1, hagen2}
starts with the Berggren completeness \cite{berggren,berggren1,berggren2,berggren3,lind}. The Gamow Shell Model
has proven to be a promising tool in assessing the structure of weakly bound and unbound nuclei. 
The Berggren completeness
is a generalized completeness which treats bound and resonant states 
on equal footing. The completeness is given by a discrete sum over bound and resonant states and 
accomponied by an integral over a non-resonant continuum of scattering states with complex energy.
A complete many-body Berggren basis may then be constructed from 
discretized single-particle Berggren orbitals
where many body Slater determinants  are constructed 
from the discrete bound, resonant and non-resonant continuum orbitals.
This is the philosophy of the Gamow Shell Model 
in full analogy with the standard Shell Model 
using a harmonic oscillator basis. 

In this paper we address the problem of how to accurately calculate
for resonances in deformed fields. Further, it is discussed how to 
obtain a complete set of
states suitable for use in  scattering/reaction problems, where spectral 
representations of Green's functions are of great interest, and 
in other many-body applications.
The study of resonances in deformed fields
has so far only rarely been considered, and for special cases. 
In Ref.~\cite{michel} the solution of the angular momentum  coupled 
Schr\"odinger equation was considered for a deformed Woods-Saxon.
They diagonalized the deformed Hamiltonian using a Berggren basis 
generated from the spherical Woods-Saxon potential in position space,
and compared with  other methods such as an expansion in oscillator functions and a direct
solution of the coupled equations. In Ref.~\cite{deform1}
energy levels and conditions for bound states to become resonances and 
resonances to become bound states were investigated for an 
axially deformed Woods-Saxon potential, by solving the 
the radial Schr\"odinger equation for coupled channels with outgoing asymptotics.
However, the coupled channels method used in Ref.~\cite{deform1} 
does not easily generalize to the non-resonant continuum. This implies
that a complete Berggren basis in a deformed field is difficult to obtain, 
and all evaluated observables will become complex quantities unless
the non-resonant continuum is taken properly into account. 
In Ref.~\cite{deform2} a different approach was considered. Their aim was
to propose a method to obtain scattering wave functions in the vicinity of a 
multi-channel resonance on the real axis, 
then calculate the phase shifts, and investigate
whether a resonance condition is met. Further this method allows for 
evaluation of observables where the continuum is properly taken into 
account, and the observables become real quantities. 

In this paper we propose 
an alternative method, starting with the momentum space Schr\"odinger equation 
given in Eq.~(\ref{eq:momspace1}). In Ref.~\cite{hagen1} it was shown 
how a complete set of Berggren states may be obtained by an analytically 
continuation of the momentum space Schr\"odinger equation in the complex $k$-plane
by utilizing the Contour-Deformation-Method (CDM). By a suitable choice of 
deformed integration contour $L^+$ we demonstrated that all 
physical resonances converges extremely fast with respect to number 
of integration points. Further it was shown that for a 
particular type of contour $L^+$, stable solutions of all physical 
scattering amplitudes may be obtained by a spectral representation
of the Green's function. 
The main difference between a  momentum space approach and 
a position space approach (see e.g. Ref.~\cite{michel}), 
lies in their different discretization schemes. 
In momentum space, it is the Bessel completeness which is discretized, while
in position space it is the completeness of the one-body problem 
(for example a Woods-Saxon completeness) which is discretized.  
The obvious advantage of the momentum space approach lies in its immediate simplicity. 
Firstly, the boundary conditions are automatically built into the integral equations. 
Secondly, the discretized Schr\"odinger equation is a complex symmetric matrix 
which is easily diagonalized, and last but not least convergence is obtained by 
increasing the number of integration points.  

In sec.~\ref{sec:formalism} the 1-dimensional angular momentum coupled 
integral equations in momentum space are derived. Sec.~\ref{sec:CDM} briefly discusses how the
integral equations may be analytically continued in the complex $k$-plane by CDM, and the relevant 
equations for numerical implementations are given.
In Sec.~\ref{sec:deformed} a deformed field of Gaussian type is introduced, and 
a brief study of different axially symmetric deformations and multipoles is given.
Sec.~\ref{sec:fourier} derives the multipole components of the Gaussian potential in 
momentum space.  
Sec.~\ref{sec:results} gives results for single-particle 
resonances in the deformed Gaussian potential, and finally 
conclusions are given in Sec.~\ref{sec:conclusions}.

\section{Momentum Space Representation of the Schr\"odinger Equation.}
\label{sec:formalism}
The momentum space single-particle Sch\"odinger equation is given by,
\begin{equation}
  {\hbar^2 \over 2\mu} k^2 \psi_n({\bf k})  + 
  \int d{\bf k'}\: V({\bf k}, {\bf k'}) \psi_n({\bf k'}) = 
  E_n \psi_n({\bf k}).
  \label{eq:momspace1}
\end{equation}
Here the notation $\psi_n({\bf k}) = \langle {\bf k} \vert \psi_n\rangle $ and 
$ \langle {\bf k} \vert V \vert {\bf k}' \rangle = V({\bf k}, {\bf k'})$ has been introduced.
The potential in momentum space is thus a double Fourier-transform 
of the potential in coordinate space, i.e.
\begin{equation} 
  V ({\bf k}, {\bf k'}) = \left( {1\over 2\pi}\right)^3 \int \mathrm{d}{\bf r}\int \mathrm{d}{\bf r}'\: 
  e^{-i {\bf k\cdot r} }V({\bf r},{\bf r}') e^{i{\bf k}'\cdot{\bf r}'}.  
\end{equation}
Here it is assumed that the interaction potential 
does not contain any spin dependence. 
Instead of an differential equation in coordinate space 
(integro-differential equation for non-local potentials), the Schr\"odinger
equation has become an integral equation in momentum space. This has 
many tractable features. Firstly, most realistic 
nucleon-nucleon interactions derived from field-theory are given 
explicitly in momentum space. Secondly, the boundary conditions imposed
on the differential equation in coordinate space are automatically built into the
integral equation. And last, but not least, integral equations are easy to numerically 
implement, and convergence is obtained by just increasing the number of integration
points.
Instead of solving the three-dimensional integral equation given in Eq.~(\ref{eq:momspace1}), an 
infinite set of 1-dimensional equations can be obtained by invoking a partial wave
expansion. 
To this end the wave function $ \psi_n({\bf k}) $ is expanded in a complete set of spherical harmonics, i.e. 
\begin{equation}
  \psi_n({\bf k}) = \sum _{lm} \psi_{nlm}(k)Y_{lm} (\hat{k}), \:\:
  \psi_{nlm}(k) = \int d\hat{k} Y_{lm}^*(\hat{k})\psi_n({\bf k}).   
  \label{eq:part_wave1}
\end{equation}
By inserting Eq.~\ref{eq:part_wave1} in Eq.~(\ref{eq:momspace1}), and projecting $Y_{lm}(\hat{k})$
from the left, 
the three-dimensional Schr\"odinger Eq.~(\ref{eq:momspace1}) is reduced
to an infinite set of  1-dimensional angular momentum coupled integral equations, 
\begin{equation}
  \left( {\hbar^2 \over 2\mu} k^2 - E_{nlm}\right) \psi_{nlm}(k) =  
  -\sum_{l'm'} \int_{0}^\infty dk' {k'}^2 V_{lm, l'm'}(k,k') \psi_{nl'm'}(k'), 
  \label{eq:part_wave2}
\end{equation}
where the angular momentum projected potential takes the form,
\begin{equation}
  V_{lm, l'm'}(k,k') = \int \mathrm{d}{\hat{k}} \int \mathrm{d}{\hat{k}'}\: 
    Y_{lm}^*(\hat{k})V({\bf k}, {\bf k'})Y_{l'm'}(\hat{k}').
    \label{eq:pot1}
\end{equation}
Here $\mathrm{d}{\hat k} = \mathrm{d}\theta \sin\theta \:\mathrm{d}\varphi $.
In many cases the potential is given in position space, so it is convienient to establish 
the connection between $V_{lm, l'm'}(k,k')$ and $V_{lm, l'm'}(r,r')$. Inserting 
position space completeness in Eq.~(\ref{eq:pot1}) gives
\begin{eqnarray}
  \nonumber
  V_{lm, l'm'}(k,k') = \int \mathrm{d}{\bf{r}} \int \mathrm{d}{\bf{r}'}\: 
  \int \mathrm{d}{\hat{k}} \int \mathrm{d}{\hat{k}'}\: 
  Y_{lm}^*(\hat{k})\langle {\bf k}\vert {\bf r} \rangle
  \langle{\bf r}  \vert V \vert {\bf r}' \rangle
  \langle {\bf r'}\vert {\bf k}' \rangle Y_{lm}(\hat{k}') = \\
  \int \mathrm{d}{\bf{r}} \int \mathrm{d}{\bf{r}'}\: 
  \left\{ \int \mathrm{d}{\hat{k}}  Y_{lm}^*(\hat{k})\langle {\bf k}\vert {\bf r} \rangle \right\}
  \langle{\bf r}  \vert V \vert {\bf r}' \rangle
  \left\{ \int \mathrm{d}{\hat{k}'}\:   Y_{lm}(\hat{k}') \langle {\bf r'}\vert {\bf k}' \rangle\right\}.
  \label{eq:pot2}
\end{eqnarray}
Since the plane waves depend only on the absolute values of position and momentum, 
$ \vert {\bf k} \vert, \vert {\bf r} \vert  $,
and the angle between them, $ \theta_{kr} $, they may be expanded in terms of bipolar harmonics of 
zero rank \cite{biedenharn}, i.e.  
\begin{equation} 
  e^{i {\bf k}\cdot {\bf r}} = 4\pi \sum_{l=0}^{\infty} i^l j_l(kr)\left( Y_l(\hat{k}) \cdot Y_l(\hat{r}) \right)
  = \sum_{l=0}^{\infty} (2l+1)i^l j_l(kr) P_l(\cos \theta_{kr}). 
\end{equation}
The addition theorem for spherical harmonics has been used in order to write
the expansion in terms of Legendre polynomials. The spherical Bessel functions, $j_l(z)$,  
are given in terms of Bessel functions of the first kind with half integer orders \cite{gradshteyn,stegun},  
\[
j_l(z) = \sqrt{\pi \over 2 z} J_{l+1/2}(z).  
\]
Inserting the plane-wave expansion
into the brackets of Eq.~(\ref{eq:pot2}) yields, 
\begin{eqnarray}
  \nonumber
  \int \mathrm{d}{\hat{k}}  Y_{lm}^*(\hat{k})\langle {\bf k}\vert {\bf r} \rangle & = &  
  \left( {1\over 2\pi} \right) ^{3/2}4\pi i^{-l} j_l(kr) Y_{lm}^*(\hat{r}), \\  
  \nonumber
  \int \mathrm{d}{\hat{k}'}\:   Y_{lm}(\hat{k}') \langle {\bf r'}\vert {\bf k}' \rangle & = &  
  \left( {1\over 2\pi} \right) ^{3/2}4\pi i^{l'} j_{l'}(k'r') Y_{l'm'}(\hat{r}). 
\end{eqnarray}
The connection between the momentum- and position space angular momentum 
projected potentials is then given by, 
\begin{equation}
  V_{lm, l'm'}(k,k') = {2 \over \pi} i^{l' -l}\int_0^\infty dr\: r^2 \int_0^\infty dr'\: {r'}^2 
  j_l(kr) V_{lm,l'm'}(r,r') j_{l'}(k'r'),
  \label{eq:pot3}
\end{equation}
which is known as a double Fourier-Bessel transform. The position space angular 
momentum projected potential is given by,
\begin{equation}
  V_{lm, l'm'}(r,r') = \int \mathrm{d}{\hat{r}} \int \mathrm{d}{\hat{r}'}\: 
  Y_{lm}^*(\hat{r})V({\bf r}, {\bf r'})Y_{l'm'}(\hat{r}').
  \label{eq:pot4}
\end{equation}
No assumptions of locality/non-locality or deformation of the interaction has so far been made, 
and the result in Eq.~(\ref{eq:pot3}) is general. In position space the Schr\"odinger equation 
takes form of an integro-differential equation in case of a non-local interaction. 
In momentum space the Schr\"odinger equation is an ordinary integral equation of the Fredholm type, 
see Eq.~(\ref{eq:part_wave2}). This is a further advantage of the momentum space approach as compared to 
the standard position space approach.  

\section{Analytic Continuation of the Momentum Space Schr\"odinger Equation by CDM.}
\label{sec:CDM}
In Ref.~\cite{hagen1} we discussed and outlined a method which analytically continues the
momentum space Schr\"odinger equation through the unitarity cut onto the second 
Riemann sheet of the complex energy plane. The method is based on deforming the integration
contour and is known as the \emph{contour deformation (distortion) method} (CDM).
As shown in Refs~\cite{hagen1,hagen2}, CDM allows for accurate calculation of a complete set of 
single-particle states, involving all kinds of poles of the scattering matrix. However, in 
Refs.~\cite{hagen1,hagen2} only spherically symmetric fields were considered. Here we wish to 
generalize the method to deformed fields, and therefore we write down the relevant equations for the 
most general case. The rules for analytic continuation of integral equation with 
general integral kernels are not outlined here, since they are the same as for 
spherically symmetric fields. We refer the reader to Ref.~\cite{hagen1} for further details 
on analytically continuation of integral equations and CDM.

The 1-dimensional coupled integral equations given in Eq.~(\ref{eq:part_wave2}) are
analytically continued from the physical to the non-physical energy sheet by 
distorting the integration contour. 
Choosing a suitable inversion symmetric 
contour $L^+$, as discussed in Ref.~\cite{hagen1}, we end up with the 
analytically continued coupled integral equations,
\begin{equation}
  \left( {\hbar^2 \over 2\mu } k^2 - E_{nlm}\right) \psi_{nlm}(k) =  
  -\sum_{l'm'} \int_{L^+} dk' {k'}^2 V_{lm, l'm'}(k,k') \psi_{nl'm'}(k'). 
  \label{eq:part_wave3}
\end{equation}
Here both $k$ and $k'$ are defined on an inversion symmetric contour $L^+$ in the lower
half complex $k$-plane, resulting in  a closed integral equation. 
The index $n$ represents a bound or resonant state.
The eigenfunctions constitute a complete bi-orthogonal set, 
normalized according to the Berggren metric \cite{berggren,berggren1,berggren2,berggren3,lind,hagen1,hagen2}. 
In solving Eq.~(\ref{eq:part_wave3}) numerically, we choose a set of 
$N$ grid points in $k-$space by some quadrature rule, for example Gauss-Legendre.
The integral is then discretized by $\int dk \rightarrow \sum_{i=1}^N w_i $.
On the chosen grid Eq.~(\ref{eq:part_wave3}) takes a complex symmetric form for 
bound, resonant and non-resonant continuum states $n$
\begin{equation}
  {\hbar^2 \over 2\mu} k_i^2 {\psi}_{n l m}(i) + 
  \sum_j^N \sum_{l'm'}\sqrt{w_i w_j} k_i k_j V_{lm,l'm'}(k_i,k_j) {\psi}_{n l'm' }(j) 
  =  E_{n l} {\psi}_{n l m}(i).
  \label{eq:momentum_space3}
\end{equation}
Changing from a continuous to a discrete plane-wave basis, it becomes transparent that
the coordinate wave function is an expansion in a basis of spherical-Bessel functions
 \begin{equation}
   \phi_{n lm}(r)  = \sqrt{ 2\over \pi}\sum_{i=1}^N \sqrt{w_i}k_i j_l(k_ir) {\psi}_{n lm}(i),
   \label{eq:wave_rad}
\end{equation}  
where ${\psi}_{n l m}(i)$ are the expansion coefficients.
Defining the functions 
\begin{equation}
  f_{l}(k_i r) = \sqrt{2 \over \pi} \sqrt{w_i} k_i j_l(k_i r),
\end{equation}
and using the discrete representation of the Dirac-delta function 
\begin{equation}
  \delta (k-k') \rightarrow {\delta _{k_i,k_j}\over \sqrt{w_i w_j}},
\end{equation}
we get the expansion 
\begin{equation}
   \phi_{n l m}(r)  = \sum_{i=1}^N {\psi}_{n l m}(i) f_l(k_ir), 
\end{equation}  
where it is easily seen that the functions $f_l(k_i r) $ are orthogonal for different $k_i$ 
and normalized
\begin{equation}
  \int dr \: r^2 f_l(k_i r)f_l(k_jr )= \delta_{k_i, k_j},
\end{equation}
$\delta_{k_i, k_j} $ being the Kronecker delta.
The complete and discrete 
set of single-particle orbits defined by this contour will then include 
the pole states, i.e., anti-bound, bound and resonant states, 
and the discretized complex continuum states defined at
each point on the contour. 

\section{Deformed field of Gaussian type.}
\label{sec:deformed}

\begin{figure}[hbtp]
  \begin{center}
    \resizebox{8cm}{8cm}
	      {\epsfig{file=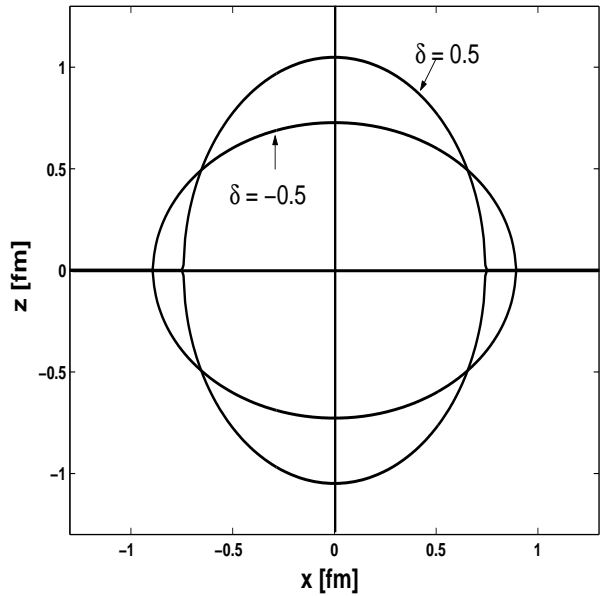}} 
  \end{center}
  \caption{Plot of the ellipsoidal isocurves $V(r,\theta) = 0.5 $ of the deformed potential for deformation
    $\delta = \pm 0.5 $ with potential strength $V_0 = 1$ in the $x,z$-plane. 
    For the spherically
    symmetric potential a  shape parameter $\alpha_0 = 1 $ was chosen.}
  \label{fig:deform1}
\end{figure}

We consider an axially symmetric deformed Gaussian potential with no spin and 
tensor components. In polar coordinates it is given as 
\begin{equation} 
  V(r,\theta) = V_0 \exp\left( -r^2( \alpha \cos^2\theta +\beta \sin^2\theta ) \right), 
  \label{eq:sph1}
\end{equation}  
or in Cartesian coordinates, 
\begin{equation} 
  V(x,y,z) = V_0 \exp\left( -\beta (x^2 + y^2)\right) \exp(-\alpha z^2), 
  \label{eq:cartesian1} 
\end{equation}  
here $V_0$ is the strength of the potential and $\alpha $ and $\beta $ are shape
parameters. Here $z$ 
is the symmetry axis, and the potential is reflection symmetric in the $x,y$-plane.
In the case $ \alpha = \beta $ 
the potential is just a spherical 
Gaussian potential. In the case $ \alpha > \beta $ 
the potential field is 
contracted along the $z$-axis, and defines an \emph{oblate} 
shape. In the case $\alpha < \beta $ 
the potential field is stretched out along the $z$-axis, and defines a \emph{prolate} shape. 
Defining a deformation parameter $\delta$ by
\begin{equation}
  \delta = 1 - { \alpha \over \beta }, 
\end{equation}
Eq.~(\ref{eq:sph1}) can be written in the form, 
\begin{equation}
  V(r,\theta;\: \beta, \delta ) = V_0 \exp (-\beta r^2) \exp( \beta \delta r^2 \cos^2\theta) = 
  V(r; \: \beta) D(r,\theta;\: \beta, \delta ).
  \label{eq:deform1}
\end{equation}
Here $V(r; \: \beta) $ 
is a spherically symmetric formfactor and $ D(r,\theta;\: \beta, \delta ) $ 
a deformation formfactor, $D=1$ for $\delta = 0$ i.e. $\alpha = \beta$.
We require that the volume of the central potential, with the shape parameter
$\alpha_0 = \alpha = \beta $, is equal to the volume of the axially symmetric
deformed ellipsoidal potential. This implies that the shape 
parameters of the non-central and central Gaussian potential satisfy the 
following relation,
\begin{equation}
  \alpha \beta^2 = \alpha_0^3, 
\end{equation} 
and the deformation parameter $\delta $ may be expressed in terms of 
$\alpha_0 $ and $\beta $ by
\begin{equation}
\delta = 1 - \left( {\alpha_0 \over \beta }\right)^3.
\end{equation}
Fig.~\ref{fig:deform1} shows plots of the isocurves $V(r,\theta) = 0.5 $ in 
the $x,z$-plane for the deformation parameters $\delta = \pm 0.5$.
With potential parameters $\alpha_0 = 1$ 
and $V_0 = 1$ for the spherically symmetric potential,
$\delta = 0.5$ 
gives 
the shape parameters  $ \alpha = 2^{-2/3}$ 
and $\beta = 2^{1/3}$ for the deformed potential,
and $\delta = -0.5 $ 
gives 
the parameters $ \alpha = (3/2)^{2/3}$ and $\beta = (2/3)^{1/3}$, respectively.
It is
seen that $ \delta = 0.5 $ 
corresponds to an \emph{prolate} shape, for $\delta = -0.5 $ 
the potential takes a \emph{oblate}
shape, the symmetry axis being the vertical $z$-axis. 
In order to assess the shape structure in more detail, it is
instructive to study the multipole components of the potential.    
An axially symmetric potential may be expanded in terms
of Legendre polynomials, i.e.  
\begin{equation}
  V(r,\theta) = \sum_\lambda V_\lambda(r) P_\lambda ( \cos\theta). 
  \label{eq:multipoles}
\end{equation}
Using the orthonormality properties of the Legendre polynomials, 
the multipole components are given by the  integrals 
\begin{equation}
  V_\lambda(r) = { (2\lambda + 1) \over 2} V_0 \exp (-\beta r^2) \int_{-1}^1 d\eta \:
  \exp( \beta \delta r^2 \eta^2 ) P_\lambda(\eta), 
\end{equation}
where $\eta = \cos(\theta) $. 
Here it is explicitly seen for our reflection symmetric potential,
that only \emph{even} multipoles give non-vanishing 
contributions, since the Legendre polynomials have the 
property 
\[
 P_\lambda(-\eta) = (-1)^\lambda P_\lambda(\eta),
\]
and the potential is an even function in $\eta$. 
\begin{figure}[hbtp]
  \begin{center}
    \resizebox{10cm}{8cm}{\epsfig{file=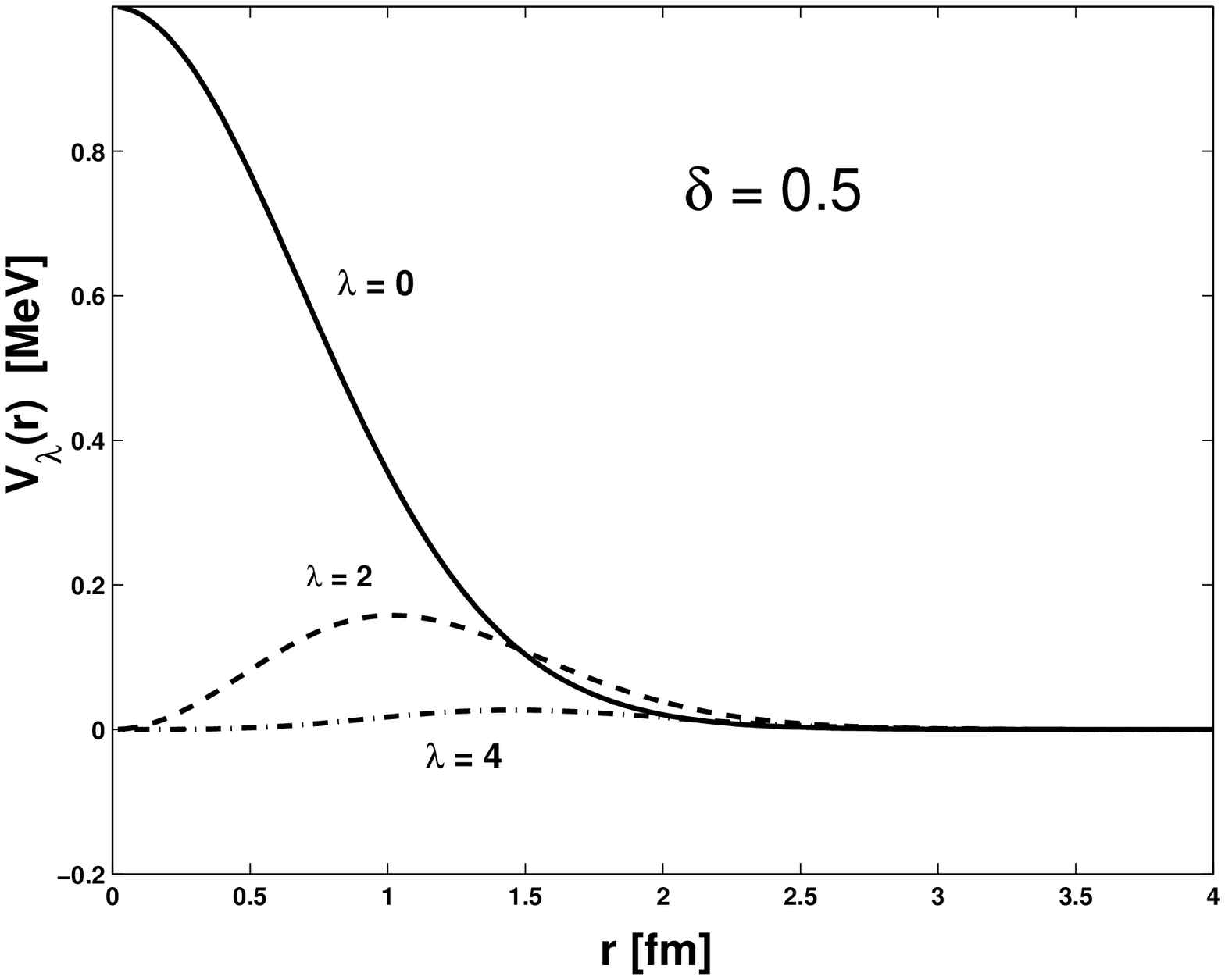}}
  \end{center}
  \caption{Plot of $\lambda = 0,2,4$ multipoles of the Gaussian potential 
    with deformation parameter $\delta = 0.5$ and $\alpha_0 = 1 $.}
\label{fig:multi1}
\end{figure}

\begin{figure}[hbtp]
  \begin{center}
    \resizebox{10cm}{8cm}{\epsfig{file=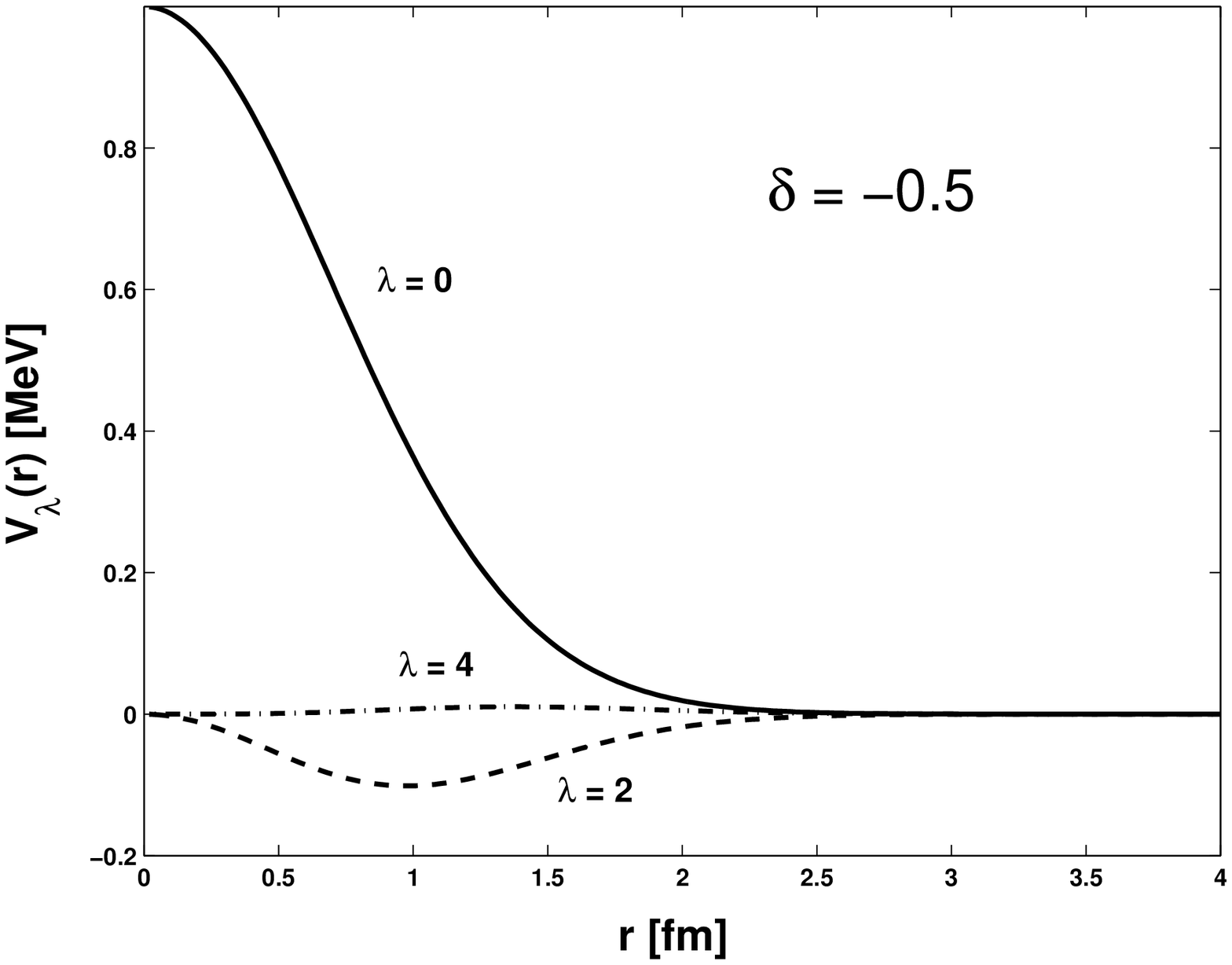}}
  \end{center}
  \caption{Plot of $\lambda = 0,2,4$ multipoles of the Gaussian potential 
    with deformation parameter  $\delta = -0.5$ and $\alpha_0 = 1 $.}
\label{fig:multi2}
\end{figure}
The monopole part of the deformed Gaussian potential may be calculated 
analytically,
\begin{equation}
  V_{\lambda=0}(r) = { 1 \over 2} V_0 \exp (-\beta r^2) \int_{-1}^1 d\eta \:
  \exp( \beta \delta r^2 \eta^2 ) = { 1 \over 2} V_0 \exp (-\beta r^2) D_0(r),
\end{equation}
where 
\begin{equation}
D_0(r) = {1\over 2\sqrt{\tau}}\gamma(1/2,\tau), \:\: \tau = -\beta \delta r^2.
\end{equation}
Here $\gamma(1/2,\tau) $ is the incomplete gamma function, see e.g. 
Ref.~\cite{boas}.
Figs.~\ref{fig:multi1} and ~\ref{fig:multi2} give plots of the 
$\lambda = 0,2,4$ multipoles of the Gaussian potential  
with deformation parameters $\delta = \pm 0.5$ and the potential parameters
$\alpha_0 = 1$ and $V_0 =1$. 
It is seen that the radial monopole 
distributions are more or less identical for $\delta = 0.5 $ and $\delta = -0.5$. 
Further it is seen that the deformed Gaussian potential 
is nearly a pure quadrupole deformation, since the $\lambda =4$ 
multipole is almost vanishing in both cases. This may be understood
from considering the exponent of the deformed formfactor in Eq.~(\ref{eq:deform1}),
which can be rewritten in terms of the $ Y_{20}(\hat{r}) $ spherical harmonic.

\section{Multipole Components in Momentum Space.}
\label{sec:fourier}
Having discussed the shape and multipoles of the deformed Gaussian potential, 
we now turn to the actual solution of the Schr\"odinger equation for 
this potential. We wish to solve the partial wave decomposed 
momentum space Schr\"odinger equation given in Eq.~(\ref{eq:momspace1}), and therefore
need the Gaussian deformed potential in a partial wave decomposed, momentum representation.
The Fourier transformation of the deformed 
Gaussian potential in Eq.~(\ref{eq:cartesian1})
is, 
\begin{eqnarray}
  \nonumber
  V(q_x,q_y,q_z) & = &  {V_0 \over (2\pi)^3 }\int dx \:dy \:dz\: 
  \exp\left( i( q_x x +q_y y +q_z z) \right) 
  \exp\left( -\beta (x^2 + y^2) -\alpha z^2 \right) \\ 
  &  = &{V_0 \over 8 \pi^{3/2} \beta \alpha^{1/2} }
  \exp\left( -{1\over 4 \beta}(q_x^2 +q_y^2) \right)   
  \exp\left( -{1\over 4 \alpha} q_z^2 \right),
\end{eqnarray}
where $q_i= k_i - k_i',\: i = x,y,z$. In terms of spherical momentum space
coordinates $k,\theta, \varphi $  
the potential takes the form, 
\begin{eqnarray}
  \nonumber
  V({\bf k},{\bf k}')&  = & {V_0 \over 8 \pi^{3/2} \beta \alpha^{1/2} }
  \exp\left( -{1\over 4 \beta} ( k^2 \sin^2\theta + {k'}^2\sin^2\theta')
  -{1\over 4 \alpha} ( k \cos\theta - {k'}\cos\theta')^2\right) \\
  & \times & \exp\left( {1\over 2 \beta } kk'\sin\theta\sin\theta' \cos\left( \varphi - \varphi'\right) \right).
  \label{eq:sph2}
\end{eqnarray} 
Due to axial symmetry the  dependence of the potential 
on the azimuthal angles $\varphi, \varphi' $ is only on the difference 
$ \omega = \varphi - \varphi'$. The potential may therefore be expanded in 
a complete set of harmonics, i.e. 
\begin{equation}
  V( {\bf k}, {\bf k'}) = \sum_{\mu = -\infty}^\infty V_{\mu}(\tilde{k}, \tilde{k}') \exp(i \mu \omega), 
  \label{eq:pot_exp1}
\end{equation}
here $\tilde{k} = (k,\theta) $. 
The harmonics $\exp ( i\mu \omega )$ obey the orthogonality relation 
\begin{equation}
  \int_{-\pi}^\pi d\omega \: \exp (-i\mu \omega) \exp( i\mu'\omega) = 2\pi \delta_{\mu,\mu'},
\end{equation}
the $\mu$'th harmonic of the potential is therefore given by the integral
\begin{equation}
  V_\mu( \tilde{k},\tilde{k}') = 
    {1 \over 2\pi} \int_{-\pi}^\pi d\omega \:\exp(-i\mu \omega) V({\bf k},{\bf k}').
\end{equation}
From Eq.~(\ref{eq:sph2}) it is seen, that for the integral over $\omega $, 
we have to consider the following integral, 
\begin{equation}
  I_{\mu}(y) = {1 \over 2\pi} \int_{-\pi}^\pi d\omega \:\exp(-i\mu \omega) 
  \exp( y \cos\omega ) = {1\over \pi} \int_0^\pi d\omega \: \cos( \mu \omega) \exp(y \cos\omega ), 
  \label{eq:bessel1}
\end{equation}
where we have introduced the variable 
\[
y = {1\over 2\beta } kk'\sin\theta\sin\theta'.
\]
The integral in Eq.~(\ref{eq:bessel1}) is just the definition of the modified Bessel 
function of the 1'st kind (see e.g. \cite{stegun}). 
The $\mu$'th harmonic of the potential is thus of analytic form, and given by, 
\begin{eqnarray}
  \nonumber
  V_\mu( \tilde{k},\tilde{k}') = {V_0 \over 8 \pi^{3/2} \beta \alpha^{1/2} } \times  \\
  \exp\left( -{1\over 4 \beta} ( k\sin\theta - {k'}\sin\theta')^2
  -{1\over 4 \alpha} ( k \cos\theta - {k'}\cos\theta')^2\right) 
  \exp(-y) I_{\mu}(y).
\end{eqnarray}
Inserting the expansion of the potential given in Eq.~(\ref{eq:pot_exp1}) into 
the momentum space Schr\"odinger Eq.~(\ref{eq:momspace1}) and projecting 
the equation on the harmonics $\exp ( i\mu \omega )$, the three-dimensional 
integral equation has been reduced to an infinite set of two-dimensional 
integral equations. The $\mu$'th integral equation is easily solved as
a matrix diagonalization problem with dimension $N_r\times N_\theta $. 
Here $N_r$ is the number of integration points for the radial integral 
and $N_\theta$ the number of integration points for the angle integral. However, the
Schr\"odinger equation can be further reduced to a coupled set of one-dimensional
integral equations by projecting on spherical harmonics (see Eq.~(\ref{eq:part_wave2})). 
The angular momentum projected potential in Eq.~(\ref{eq:sph2}) then  takes the form, 
\begin{eqnarray}
  \nonumber
  V_{lm, l'm'}(k,k') = \int \mathrm{d}{\hat{k}} \int \mathrm{d}{\hat{k}'}\: 
  Y_{lm}^*(\hat{k})\left\{ \sum_{\mu = -\infty}^\infty V_{\mu}(\tilde{k}, \tilde{k}') 
  \exp(i \mu \omega)\right\}   Y_{l'm'}(\hat{k}') \\
  = 2\pi \int_0^\pi d\theta \: \sin\theta \int_0^\pi d\theta' \: \sin\theta' 
  \bar{P}_{lm}(\cos\theta) V_{m}(\tilde{k}, \tilde{k}')  \bar{P}_{l'm}(\cos\theta') \:\delta_{m,m'},
  \label{eq:pot_exp2}
\end{eqnarray}
where $ \bar{P}_{lm}(x)$ are the normalized associated Legendre polynomial, 
\begin{equation}
  \bar{P}_{lm}(x) = \left\{ { 2l + 1 (l-m)! \over 2 (l+m)! }\right\}^{1/2}P_{lm}(x).
\end{equation}

In our calculations we start with the angular momentum projected potential 
given in Eq.~(\ref{eq:pot_exp2}).
In numerical calculations the multipole expansion in angular momentum $l$ has to 
be truncated at some $l_{\mathrm{max}}$ and the Hamiltonian matrix  to be diagonalized is,
\begin{equation}
  H^{ m \pi} = 
  \left[ \begin{array}{ccc} 
      H^{m\pi}(l_1,l_1) & \ldots &  H^{m \pi}(l_1, l_{\mathrm{max}}) \\
      \vdots &  & \vdots \\
      H^{m \pi}(l_{\mathrm{max}},l_1) & \ldots &  H^{m \pi}(l_{\mathrm{max}}, l_{\mathrm{max}})
    \end{array}\right]. 
  \label{eq:partial1}
\end{equation} 
Here $ m $ and $  \pi $ are the angular momentum projection and parity, respectively,  
which are good quantum numbers in case of axial symmetry. The allowed values of $l$ are
even and odd for positive and negative parity states, respectively.
Each submatrix $ H^{m \pi}(l,l') $
in Eq.~(\ref{eq:partial1}) has matrix elements given by, 
\begin{equation}
 H^{m \pi}_{i,j}(l,l') = {\hbar^2 \over 2\mu} k_i^2 \delta_{i,j}\delta_{l,l'} + 
 \sqrt{w_i w_j} k_i k_j V_{lm,l'm}(k_i,k_j).
\end{equation}
The rank of each submatrix  $ H^{m \pi}(l,l') $ are determined by the 
total number of integration points used in the discretization of the integration contour 
$L^+$ in the coupled momentum space Schr\"odinger equation, given in Eq.~(\ref{eq:part_wave3}).
The results  reported in this work, used the same number of integration points for 
each coupled equation, so the total rank of the matrix to be diagonalized is 
$  N \times N_l $, where $N$ is the total number of integration points and 
$ N_l$ the total number of angular momentum coupled integral equations
given in Eq.~(\ref{eq:part_wave3}). Diagonalizing the complex symmetric matrix (\ref{eq:partial1}),    
we obtain a complete set of states within the chosen discretization space. The basis 
may be utilized in different spectral represenations used in scattering theory
or in Gamow-Shell-Model calculations involving deformed fields.

\begin{figure}
  \begin{center}
    \resizebox{10cm}{8cm}{\epsfig{file=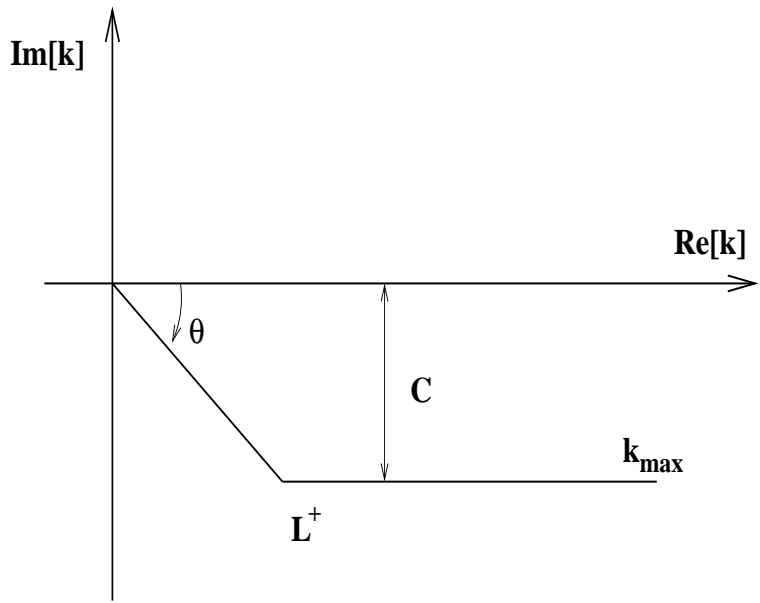}}
  \end{center}
  \caption{Sketch of the contour $L^+$ in the complex $k$-plane. The contour 
    is specified by a rotation angle $\theta $ and a translation $C$ in the 
    fourth quadrant of the complex $k$-plane.}
    \label{fig:contour1}
\end{figure}

In all calculations reported below, we used a discretized contour $L^+$ 
defined by a rotation $\theta $ and a translation $C$ in the complex $k$-plane, 
see Fig.~\ref{fig:contour1}. 
In Ref.~\cite{hagen1} it was shown that this type of contour 
allows for a stable numerical solutions of the scattering amplitude 
(or $t-$matrix), by using a spectral representation of the Green's function. 
In physical scattering, the energy is given along the real axis. 
By defining a basis with continuum energies given along the contour
depicted in Fig.~\ref{fig:contour1}, the problem with poles of the Green's function
are eliminated, see Ref.~\cite{hagen1} for more details.
The total number of integration points is given by $N = N_R + N_T$, where
$N_R$ is the number of points along the rotated line defined by $\theta$ and 
$N_T$ is the number of points along the translated line defined by $C$. 
In numerical calculations $ k_{\mathrm{max}}$ (see Fig.~\ref{fig:contour1})
should be chosen large enough, so that the calculated wave functions and energies 
do not change with increasing $ k_{\mathrm{max}}$. In our calculations 
we used $ k_{\mathrm{max}} = 6 \mathrm{fm}^{-1}$.

\section{Formation of single-particle resonances in a deformed Gaussian potential.}
\label{sec:results}
As a model study we consider the Gaussian potential given in Eq.~(\ref{eq:sph1}), 
which in the spherically 
symmetric case reproduces the $J^\pi={3/2^{-}_1} $ resonance in $^5$He. 
The  $J^\pi={3/2^{-}_1}$ resonance, to be associated with the single-particle orbit 
$p_{3/2}$, is experimentally 
known to have a width of $\Gamma \approx 0.60$ MeV.

In our calculations we used the 
following parameters for the spherically symmetric Gaussian given in 
Eq.~(\ref{eq:sph1}),
\begin{equation}
  V_0 = -53.5 \mathrm{MeV}, \:\: \alpha_0 = 0.188 \mathrm{fm}^{-2}.
\end{equation}
As a check of our momentum space approach, we 
compared our results for the spherical limit with the results 
obtained with the computer program GAMOW \cite{gamow_code}. For the spherical Gaussian potential, 
GAMOW gives a bound state
for the $l^\pi = 0^+$ channel with energy $E = -14.9044  $MeV, 
and a resonance for the $l^\pi = 1^-$ channel with energy 
$E = 0.7268  -0.3096i $MeV. Here the nucleon spin $s=1/2$ 
is neglected
since the energy levels are degenerate for $ j = l \pm 1/2 $, which 
follows from the spin independence of the Gaussian potential. Our momentum space 
calculations were able to exactly reproduce these results with a total number 
of discretization points $N = 25$. This comparison of two different methods,
provided us with a check of our codes and our derivation of the momentum space equations.

In order to obtain converged results for the deformed case, we investigate the convergence 
with respect to total number angular momentum coupled equations,
see Eq.~(\ref{eq:partial1}), and with respect to the total number 
of integration points used in discretization of each 
coupled integral equation given in Eq.(\ref{eq:part_wave3}).
First, convergence with respect to total number 
angular momentum coupled equations in Eq.~(\ref{eq:partial1}) is considered. 
We fixed the number of integration points to $N = N_R + N_T = 20 + 30 = 50$, 
this is large enough to ensure convergence with respect to number
of discretization points. For bound states we used a real contour $L^+$, i.e. 
$\theta = 0$ and the real $k-$axis was discretized with 50 points. 
In the case of resonant states we used a complex contour $L^+$ 
defined by a rotation $\theta = \pi/4 $ and a translation
$C = \sin(\pi /4)\times 0.4 \mathrm{fm}^{-1} \sim -0.29\mathrm{fm}^{-1}$ in the complex $k$-plane
(see Fig.~\ref{fig:contour1}).

\begin{figure}
  \begin{center}
    \resizebox{10cm}{8cm}{\epsfig{file=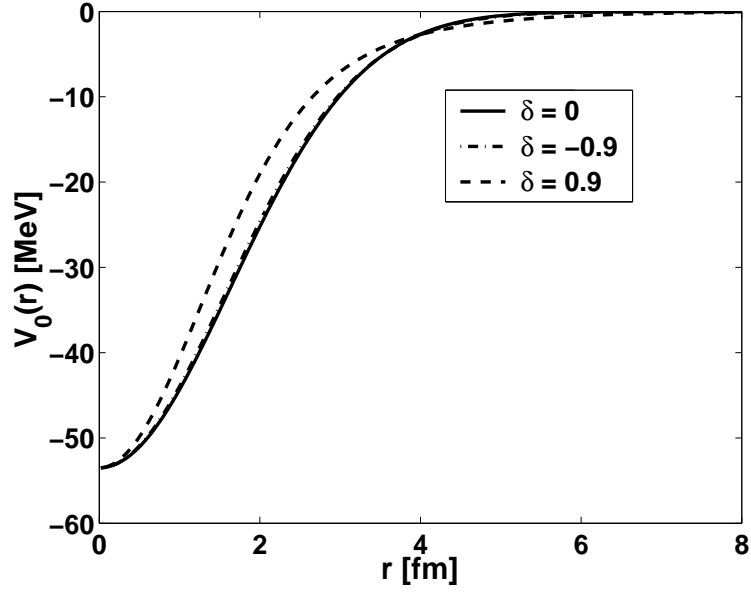}}
  \end{center}
  \caption{Plot of the monopole part  of the Gaussian potential 
    with deformation parameters $\delta = 0$ and $\delta = \pm 0.9$.}
  \label{fig:monopole}
\end{figure}

Table~\ref{tab:deform1} gives the convergence of the $m^\pi = 0^+$ 
ground state energy, for deformation parameters $ \delta = \pm 0.9$. 
For $ \delta = -0.9$, convergence is quickly reached, with  $l_{\mathrm{max}} = 4 $.
For $ \delta =  0.9$, convergence is considerably slower. 
It is seen that
the deformation $\delta = 0.9$ affects the bound state most, and 
the ground state becomes less bound $ E = -12.1 $MeV, 
for the prolate deformation. 
On the other hand, the oblate deformation $\delta = -0.9$ has little effect
on the ground state energy $E = -14.7$MeV.  
\begin{table}[htbp]
  \begin{center}
  \begin{tabular}{ccccc}
    \hline
    \multicolumn{1}{c}{} & \multicolumn{2}{c}{$\delta = 0.9 $} 
    & \multicolumn{2}{c}{$\delta = -0.9$} \\
    \hline
    \multicolumn{1}{c}{$l_{\mathrm{max}}$} &
      \multicolumn{1}{c}{Re[E]} & \multicolumn{1}{c}{Im[E]} &
      \multicolumn{1}{c}{Re[E]} & \multicolumn{1}{c}{Im[E]} \\
    \hline
    0 &  -10.5843 & 0.&  -14.4816 & 0.\\
    2 &  -11.9041 & 0.&  -14.6533 & 0.\\
    4 &  -12.0741 & 0.&  -14.6551 & 0.\\
    6 &  -12.0953 & 0.&  -14.6551 & 0.\\
    8 &  -12.0979 & 0.&  -14.6551 & 0.\\ 
    10 & -12.0983 & 0.&  -14.6551 & 0.\\
    \hline
  \end{tabular}
  \caption{Convergence of groundstate, $m^\pi = 0^+$, for deformation
    parameters  $\delta = \pm 0.9$ as the number of partial waves increases.
    In the spherically symmetric case $\delta = 0$ the  $l^\pi=0^+$ Gaussian potential 
    supports a bound state at energy $ E = -14.9044 $MeV. }
  \label{tab:deform1}
  \end{center}
\end{table}
This may be understood by considering the monopole term of the potential, 
which is the main component in the multipole expansion in Eq.~(\ref{eq:multipoles}).
In Fig.~\ref{fig:monopole} a plot of the monopole part of the Gaussian 
potential with deformation parameters $\delta = \pm 0.9$ is given, together with a plot 
of the spherically symmetric potential. It is
seen that the monopole term for the $\delta = -0.9$ 
potential is more or less identical 
to the spherically symmetric potential (slightly less attractive), 
on the other hand the monopole 
term for the $\delta = 0.9 $ potential is less attractive for $ r < 4$fm, but
more attractive at large distances $r>4$fm.  From this one may conclude that the 
ground state of the $\delta = -0.9$ 
potential will be more bound than for the $\delta = 0.9$ potential, since 
the ground state is deeply bound and the wave function will be
mainly located in the interior part of the potential.

The resonant orbit $l^\pi = 1^-$ in the spherically symmetric 
potential is split into two non-degenerate orbits ( $m^\pi = 0^- $ and $ m^\pi  = 1^-$) 
in the case of an axially symmetric deformation. This is a 
characteristic of axially deformed potentials.
Table~\ref{tab:deform2} gives the convergence of the $m^\pi = 0^-$ and  
$m^\pi = 1^-$ excited negative parity states in the Gaussian potential,
for deformation parameters $ \delta = \pm 0.5$. In all cases a satisfactory 
convergence is obtained with $l_{\mathrm{max}} = 5$. 
For the states 
with vanishing angular momentum projection along the $z$-axis ( $ m = 0$ ), it is
seen that for $\delta = 0.5$ ( prolate deformation ) the $ l^\pi = 1^-$ 
state
has become a bound state with energy $ E = -0.680 $MeV. 
For zero angular momentum projection, the particle moves in an orbit
along the $z$-axis. So in the case of a prolate deformation, 
where the field is stretched out along the $z$-axis, a particle moving in this
orbit will ``feel'' the field more strongly than compared with the spherically symmetric 
field, and it will become more bound. This explains also why the particle 
with $m=0$ becomes more unbound in the case of the oblate deformation $\delta = -0.5$,
see Columns 6 and 7 of Table~\ref{tab:deform2}.  
\begin{table}[htbp]
  \begin{center}
  \begin{tabular}{ccccccccc}
    \hline
    \multicolumn{1}{c}{} & \multicolumn{4}{c}{$\delta = 0.5 $} 
    & \multicolumn{4}{c}{$\delta = -0.5$} \\
    \hline
    \multicolumn{1}{c}{} & \multicolumn{2}{c}{$m^\pi = 0^- $} 
    & \multicolumn{2}{c}{$ m^\pi = 1^- $} 
    & \multicolumn{2}{c}{$m^\pi = 0^- $} 
    & \multicolumn{2}{c}{$ m^\pi = 1^- $} \\
    \hline
    \multicolumn{1}{c}{$l_{\mathrm{max}}$} &
    \multicolumn{1}{c}{Re[E]} & \multicolumn{1}{c}{Im[E]} &
    \multicolumn{1}{c}{Re[E]} & \multicolumn{1}{c}{Im[E]} &
    \multicolumn{1}{c}{Re[E]} & \multicolumn{1}{c}{Im[E]} &
    \multicolumn{1}{c}{Re[E]} & \multicolumn{1}{c}{Im[E]} \\
    \hline
    1 &   -0.5282 & 0. &  1.4865 & -1.0177 &  1.5402 & -1.0701 &  0.3815 & -0.1139 \\
    3 &   -0.6772 & 0. &  1.4419 & -0.9631 &  1.5170 & -1.0404 &  0.3602 & -0.1042 \\
    5 &   -0.6802 & 0. &  1.4410 & -0.9621 &  1.5168 & -1.0402 &  0.3601 & -0.1041 \\
    7 &   -0.6803 & 0. &  1.4410 & -0.9620 &  1.5168 & -1.0402 &  0.3601 & -0.1041 \\
    9 &   -0.6803 & 0. &  1.4410 & -0.9620 &  1.5168 & -1.0402 & 0.3601 & -0.1041  \\
    \hline
  \end{tabular}
  \caption{Convergence of the $m^\pi = 0^-$ and $m^\pi = 1^-$ energies, for deformation
    parameters  $\delta = \pm 0.5$ with increasing number of partial waves.
    In the spherically symmetric case ($\delta = 0$) the  $l^\pi=1^-$ Gaussian potential 
    supports a resonance state at energy $ E = 0.7268  -0.3096i $MeV.}
\label{tab:deform2}
\end{center}
\end{table}

\begin{table}[htbp]
  \begin{center}
  \begin{tabular}{ccccccccc}
    \hline
    \multicolumn{1}{c}{} & \multicolumn{4}{c}{$\delta = 0.5 $} 
    & \multicolumn{4}{c}{$\delta = -0.5$} \\
    \hline
    \multicolumn{1}{c}{} & \multicolumn{2}{c}{$m^\pi = 0^- $} 
    & \multicolumn{2}{c}{$ m^\pi = 1^- $} 
    & \multicolumn{2}{c}{$m^\pi = 0^- $} 
    & \multicolumn{2}{c}{$ m^\pi = 1^- $} \\
    \hline
    \multicolumn{1}{c}{$l_{\mathrm{max}}$} &
    \multicolumn{1}{c}{Re[$\psi_{l}^2$ ]} & \multicolumn{1}{c}{Im[$\psi_{l}^2$]} &
    \multicolumn{1}{c}{Re[$\psi_{l}^2$ ]} & \multicolumn{1}{c}{Im[$\psi_{l}^2$]} &
    \multicolumn{1}{c}{Re[$\psi_{l}^2$ ]} & \multicolumn{1}{c}{Im[$\psi_{l}^2$]} &
    \multicolumn{1}{c}{Re[$\psi_{l}^2$ ]} & \multicolumn{1}{c}{Im[$\psi_{l}^2$]} \\
    \hline
    1 &   0.9947 & 0. & 0.9982 & 2.31E-03& 0.9992 & 1.1E-03 & 0.9993 & 3.E-04 \\ 
    3 &   5.7E-03&0. & 1.8E-03 &-2.3E-03& 8.E-04&-1.1E-03& 7.E-04 &-3.E-04 \\
    5 &   5.E-05 & 0. &2.E-05  &-2.E-05 & 3.E-06&-3.E-06& 2.E-06 &-9.E-07 \\
    7 &   6.E-07 & 0. &3.E-07  &-2.E-07 & 2.E-08&-1.E-08& 9.E-09 &-4.E-09 \\
    9 &   8.E-09 & 0. &4.E-09  &-3.E-09 & 9.E-11&-7.E-11& 4.E-11 &-2.E-11 \\
    \hline
  \end{tabular}
  \caption{Convergence of the $m^\pi = 0^-$ and $m^\pi = 1^-$ squared amplitudes 
    of the wave functions for each partial wave $l$, for deformation
    parameters  $\delta = \pm 0.5$.}
\label{tab:deform3}
\end{center}
\end{table}
For the $m=1$ case the opposite
takes place. In the case of $\delta = 0.5$ the particle becomes more unbound, 
while for $\delta = -0.5$ the particle becomes more bound. By considering 
the dipole ($l=1$) term of the wave function, the particle moves in 
an orbit 
making $ \pi/4$ 
degrees with the $z$-axis. From this
it may be understood that the particle gains more binding in the case of 
an oblate deformation $\delta = -0.5$ and becomes more un-physical in the 
opposite case $\delta = 0.5$ 
( see columns 4,5,8 and 9 of table~\ref{tab:deform2}).

In table~\ref{tab:deform3} the squared amplitudes of the wave functions are given
for each partial wave $l$. It is seen that in all cases that the 
squared amplitudes for the $l=1$ component of the total wave function, 
is nearly equal to the norm of the total wave function, while
all other partial wave amplitudes are vanishing small. In this sense one may 
say that the orbital angular momentum is approximately a ``good'' quantum number. 

Having investigated the convergence with respect to number 
of angular momentum coupled equations, we now consider convergence with 
respect to number of discretization points along the contour $L^+$, 
for the negative parity states. 
For the energies in Tables~\ref{tab:deform2} 
and \ref{tab:deform3}, we reached satisfactory convergence with 
$l_{\mathrm{max}} = 7 $. In considering convergence with respect to 
integration points, we then fix the maximum number of coupled equation to 
$l_{\mathrm{max}} = 7 $. Table~\ref{tab:deform4} reports the 
convergence of the odd parity energies in the deformed Gaussian
potential given in Eq.~(\ref{eq:sph2}), with deformation parameters $\delta = \pm 0.5$.  
\begin{table}[htbp]
  \begin{center}
  \begin{tabular}{cccccccccc}
    \hline
    \multicolumn{2}{c}{} & \multicolumn{4}{c}{$\delta = 0.5 $} 
    & \multicolumn{4}{c}{$\delta = -0.5$} \\
    \hline
    \multicolumn{2}{c}{} & \multicolumn{2}{c}{$m^\pi = 0^- $} 
    & \multicolumn{2}{c}{$ m^\pi = 1^- $} 
    & \multicolumn{2}{c}{$m^\pi = 0^- $} 
    & \multicolumn{2}{c}{$ m^\pi = 1^- $} \\
    \hline
    \multicolumn{1}{c}{$N_R$} &
    \multicolumn{1}{c}{$N_T$} &
    \multicolumn{1}{c}{Re[E]} & \multicolumn{1}{c}{Im[E]} &
    \multicolumn{1}{c}{Re[E]} & \multicolumn{1}{c}{Im[E]} &
    \multicolumn{1}{c}{Re[E]} & \multicolumn{1}{c}{Im[E]} &
    \multicolumn{1}{c}{Re[E]} & \multicolumn{1}{c}{Im[E]} \\
    \hline
5  &  10 &  -0.6777 & 0.0000 & 1.4392 & -0.9680 & 1.5150 & -1.0472 & 0.3576 & -0.1091 \\
10 &  10 &  -0.6777 & 0.0000 & 1.4401 & -0.9656 & 1.5150 & -1.0448 & 0.3577 & -0.1091 \\
10 &  15 &  -0.6803 & 0.0000 & 1.4411 & -0.9621 & 1.5170 & -1.0403 & 0.3601 & -0.1043 \\
10 &  20 &  -0.6803 & 0.0000 & 1.4410 & -0.9620 & 1.5168 & -1.0402 & 0.3601 & -0.1041 \\
10 &  25 &  -0.6803 & 0.0000 & 1.4410 & -0.9620 & 1.5168 & -1.0402 & 0.3601 & -0.1041 \\
15 &  25 &  -0.6803 & 0.0000 & 1.4410 & -0.9620 & 1.5168 & -1.0402 & 0.3601 & -0.1041 \\
15 &  30 &  -0.6803 & 0.0000 & 1.4410 & -0.9620 & 1.5168 & -1.0402 & 0.3601 & -0.1041 \\
20 &  30 &  -0.6803 & 0.0000 & 1.4410 & -0.9620 & 1.5168 & -1.0402 & 0.3601 & -0.1041 \\
    \hline
  \end{tabular}
  \caption{Convergence of the $m^\pi = 0^-$ and $m^\pi = 1^-$ energies
    with increasing number of discretization points along 
    the contour $L^+$. Here  deformation
    parameters  $\delta = \pm 0.5$ were used, and $l_{\mathrm{max}} = 7$.}
  \label{tab:deform4}
\end{center}
\end{table}
It is seen that one obtains convergence with a total 
number of integration points given by $N=N_R + N_T = 10 + 20 = 30$. 
Note also that with $N=15$ points, we have satisfactory results. For $N=30$ and 
$N_l = 4 $ (only four coupled equations for $l_{\mathrm{max}} = 7$ due to conservation of parity)
the total dimension of the matrix in Eq.~(\ref{eq:partial1}) is $dim =120$, which is 
diagonalized extremely fast with any diagonalization routine suitable for complex symmetric 
matrices.

\begin{figure}[hbtp]
  \begin{center}
    \resizebox{10cm}{8cm}{\epsfig{file=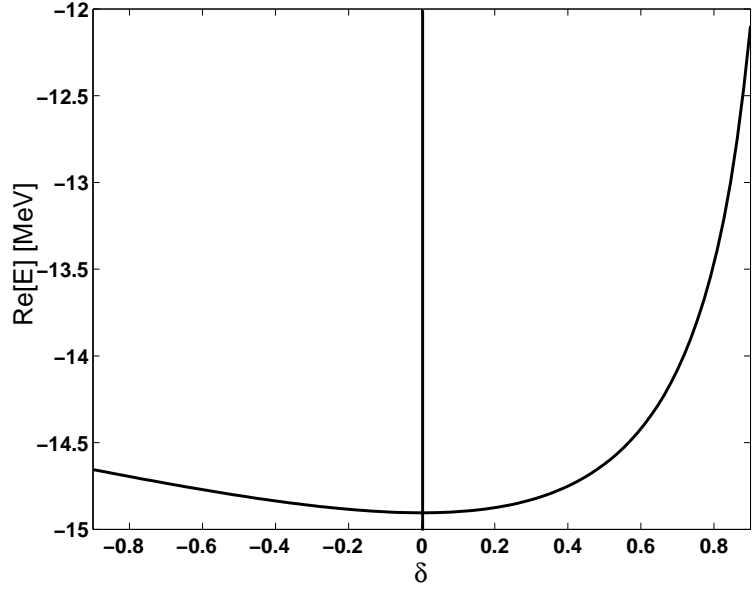}}
  \end{center}
  \caption{Bound state trajectory for the $m^\pi = 0^+$ state in the deformed 
  Gaussian potential. Energy is plotted as function of deformation parameter $\delta$.}
\label{fig:0plus}
\end{figure}

Having determined convergence properties of the odd and even parity states in the 
deformed Gaussian potential, we now study how the different states behave over a large
range of deformations. 
In figure~(\ref{fig:0plus}) a plot of the bound state energy of the $m^\pi = 0^+$ 
state is given for the deformation parameter $\delta $ 
taking values between $-0.9$ and $0.9$. 
It is seen that the position of the
bound state varies much more strongly for a prolate deformation ($\delta > 0$), than 
for an oblate deformation. 

\begin{figure}[hbtp]
  \begin{center}
    \resizebox{10cm}{8cm}{\epsfig{file=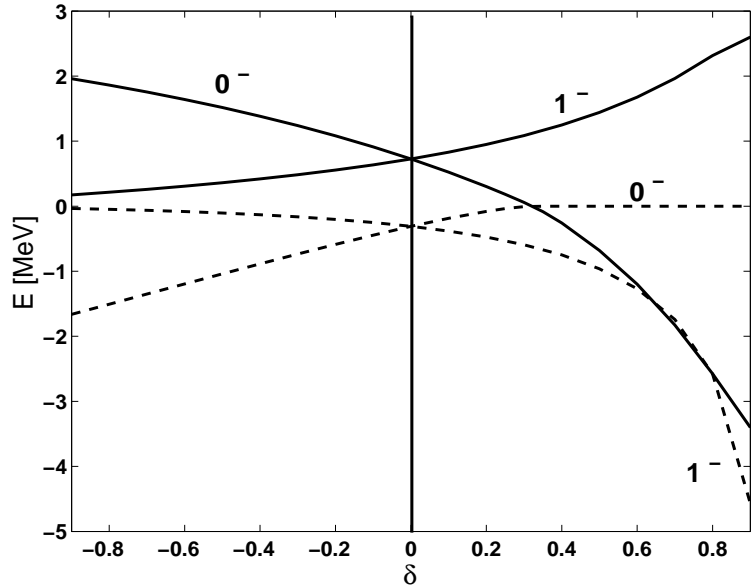}}
  \end{center}
  \caption{Real (solid lines) and imaginary parts (dashed lines) 
    of the $m^\pi = 0^-$ and the $m^\pi = 1^-$ 
    state energies in the deformed 
    Gaussian potential as the deformation parameter $\delta $ is varied between $-0.9$ 
    and $0.9$.}
  \label{fig:combined}
\end{figure}

Fig.~\ref{fig:combined} shows a plot of the real (solid lines) and imaginary part (dashed lines) 
of the $m^\pi = 0^-$ and $m^\pi = 1^-$ states for the deformation parameter $\delta $ 
taking values between $-0.9$ and $0.9$.

The value for $\delta$ in which the $m^\pi = 0^-$  resonant state becomes a bound state
is given when the real energy trajectory meets the imaginary energy 
trajectory for $ \mathrm{Re}[E]< 0$, i.e. $\mathrm{Im}[E] = \mathrm{Re}[E]$. 
Here the splitting of the $l^\pi = 1^-$ 
resonant level with respect to the angular momentum projection $m$ is clearly seen.
It is also seen that the energy of the $m^\pi = 0^-$ 
and the $ m^\pi = 1^-$ state behave in opposite manner for $\delta > 0$ and for $\delta <0$. 
The $m^\pi = 0^-$  resonance for $\delta =0$ becomes a 
bound state for $\delta > 0.3 $. For $\delta < 0$ the $m^\pi = 0^-$  
resonance moves further down in the lower half complex $k$-plane. 
On the other hand, it is seen that the  $m^\pi = 1^-$ resonance state does not become a
bound state for the values of $\delta $ considered, $\delta \in (-0.9,0.9)$. 
For $\delta \in ( -0.9,0.2) $ 
the $m^\pi = 1^-$ resonance energy displays a weak variation from the $\delta = 0$ energy, and
slowly moves towards the scattering threshold $E=0$, for $\delta \rightarrow -0.9$. 
On the other hand, as $\delta \rightarrow 0.9 $ the imaginary part of the energy
dives into the lower half complex energy plane, and the resonance
state becomes strongly unphysical.

\section{Conclusion.}
\label{sec:conclusions}
In this work, we have demonstrated that the Contour Deformation Method 
may be generalized to the case of deformed potentials. CDM is applied to the momentum space 
Schr\"odinger equation, allowing for stable and converged solutions of physical resonances. In addition 
a complete set of Berggren states is obtained, which may be used in the construction of a many-body 
Berggren basis. The most obvious advantage of this momentum space approach, as compared to its position space analog, 
is that the 
boundary conditions of all kinds of states, i.e. bound, resonant and continuum, are automatically taken 
care of since we are dealing with integral equations instead of integro-differential equations. 
The method is demonstrated for axial symmetry and a fictitious "deformed $^5$He", but may be extended to more
general deformation and applied to truly deformed halo nuclei.
Such applications as to $^8$Li, $^9$Be and $^{11}$Be will be discussed elsewhere.


\end{document}